\documentclass[11pt, a4paper, twocolumn, copyright, goog]{google}

\usepackage{xcolor}


\uselogo{} 

\title{Beyond Zero: Enterprise Security for the AI Era}

\renewcommand{\today}{2026-05-21}

\author[1]{Joseph Valente}
\author[2]{Michal Zalewski}

\affil[1]{Director of Product Management, Alphabet Security, Alphabet, Inc.}
\affil[2]{Distinguished Engineer, Alphabet Security, Alphabet, Inc.}

\begin{abstract}
The rise of autonomous AI agents and the accelerating velocity of
corporate data access are stretching the application-centric model of
zero trust security to its breaking point. This paper introduces
\emph{Beyond Zero}, a new security paradigm designed for the AI era. The
\emph{Beyond Zero} architecture performs per-resource and method access
decisions for humans and agents at machine speed. By shrinking the
trust boundary from the application level to the individual action, and
by coupling static authorization guarantees with dynamic, AI-driven
reasoning, \emph{Beyond Zero} enables a self-defending enterprise capable of
mediating thousands of human and machine decisions per second. This
paper outlines Google's vision for the future of this access model as
well a call for industry collaboration and standards development.
\newline

This is a preprint and this paper has been accepted for publication in ACM Queue. The final version of this paper may change through the editorial process. 
\end{abstract}

\begin{document}

\maketitle

\section{Introduction}\label{introduction}

Enterprise security is in the process of transitioning from human-speed
security to high-frequency, AI-mediated defense. We call this new model
\emph{Beyond Zero}, and this paper lays out the fundamental concepts for the new
architecture.

\emph{Beyond Zero} allows for a contextual, risk-based, resource-level
authorization model to run at machine-speed, securing both humans and
agents without overburdening users.

The \emph{Beyond Zero} model has a number of key features:

\begin{enumerate}
\def\labelenumi{\arabic{enumi})}
\item
  \textbf{Resource/action-based security:} Individual actions on individual
  resources are where authorization decisions are made, rather than
  provisioning access to a tool, application, or feature within an
  application. This is true no matter where those resources are
  accessible from (e.g. via front-end tooling, API, MCP, or other access
  methods).
\item
  \textbf{Blended static and dynamic security:} Granular static policies are
  applied to resource accesses in addition to fully dynamic controls
  that apply even stronger security in risky or complex scenarios. This
  allows for dynamic controls to be applied without shifting to a fully
  dynamic security model that is challenging to statically verify.
\item
  \textbf{Automatically enriched context:} The decision systems can draw on
  context about the user, what the user should be working on, what data
  the user is trying to interact with, what the user is attempting to do
  with the data, and what potential risk mitigations are available.
  These facts are always available to the decision making
  infrastructure.
\item
  \textbf{Automated in-depth investigation:} Investigations can be triggered to
  run autonomously by risk signals, and in turn can activate either
  challenges or containments that are immediately applied to the user's
  stream of accesses.
\item
  \textbf{Challenges and containments:} Challenges and containments can be
  triggered directly from the security policies, enabling accessors to
  provide additional risk information on demand.
\end{enumerate}

This model builds on the BeyondCorp model Google introduced in
2014,\footnote{BeyondCorp: A New Approach to Enterprise Security, Rory
  Ward, Betsy Beyer, ;login:, Vol. 39, No. 6 (2014), pp. 6-11.} by
extending the Zero Trust concept to the authentication and authorization
layers. The assumptions underpinning BeyondCorp---that accessors are
human, that actions occur at human speed, and that applications are the
correct boundary for trust---are no longer sufficient. Now that we are
entering an era where AI agents access data at orders of magnitude higher than the rate of humans, often reasoning about information across vast, unstructured datasets, we
 need a new model for securing data. Note that when we discuss AI agents
in this paper, we're specifically referring to AI systems that autonomously take actions on behalf of a user or team within an enterprise.

The publication of the BeyondCorp whitepaper marked both a milestone on
Google's own journey towards adopting the BeyondCorp model, as well as a
vision for how the rest of the industry could both adopt the same model
as well as build the infrastructure to help others adopt a similar
model. In a similar vein, this publication marks a milestone in our own
effort to transition to the \emph{Beyond Zero} model as well as a vision for
where we believe the industry needs to move to accelerate this
transition.
\section{The Problem: Securing data at machine
speed}\label{the-problem-securing-data-at-machine-speed}

A number of major shifts have made the transition to the \emph{Beyond Zero}
model necessary. In particular, some of the major changes we have
observed are:

\emph{Agents Driving Exponential Growth in Access Volume}

The transition from a human-centric workforce to an agentic one has
fundamentally altered the frequency of access requests. AI agents,
whether operating autonomously or as `copilots' for human users,
interact with corporate resources at a far higher rate than traditional
human activity. Our legacy infrastructure is strained by the need to
authorize tens of millions of concurrent machine-driven actions per
second.

\emph{Unprecedented Data Velocity and Scale}

It is not merely the frequency of access that has changed, but the
volume of data being consumed. Even before the adoption of agents, there
was already an exponential growth in both the volume and sensitivity of
data being accessible inside digital applications. Additionally, with
agentic workflows, we're now seeing a rise in aggregation and reasoning
across vast, unstructured datasets, leading to an even greater growth in
the volume of data. This "geometric shock" to the system means that more
sensitive intellectual property and user data are capable of being
accessed and acted upon than ever before. While AI enables this data to
be processed and acted upon more quickly, any processing would happen at
a velocity that outpaces manual oversight and after-the-fact secops
review.

\emph{Increasing Sophistication of the Threat Landscape}

In the current environment even the least sophisticated attackers have
weaponized AI to accelerate their tradecraft. These adversaries use
large language models (LLMs) to rewrite malicious code on demand, making
static detection methodologies less effective. Attackers can now afford
to be patient, testing surfaces that were previously considered to be
low value using automated agentic methods. Furthermore, the rise of
agentic infrastructure introduces net-new attack vectors, such as the
exploitation of "ambient authority," where an agent is granted the full,
often over-provisioned, permissions of its human user. This combination
of machine-speed attacks and non-deterministic agent behavior creates a
risk profile that requires a move toward dynamic, intent-based defense.
\section{The Solution: Introducing the \emph{Beyond Zero}
model}\label{the-solution-introducing-the-Beyond Zero-model}

\emph{Beyond Zero} shifts the trust boundary from the application to the action
being performed on a piece of data in real-time -- and from
after-the-fact investigation to in-the-moment evaluation and
containment. It augments BeyondCorp's foundational identity with a
"brain" capable of reasoning about the \emph{context} and \emph{intent}
of a specific request in real-time.

\begin{table*}[t]{}
\centering
    \begin{tabular}{p{0.1911\textwidth} | p{0.1831\textwidth} | p{0.54\textwidth}}
        \hline\rowcolor{gray!20}
        Feature & BeyondCorp (Legacy) & Beyond Zero (New) \\
        \hline
        Trust Boundary & Application / Tool & Individual Action / Resource   \newline             \emph{Limits the boundary of trust to an individual piece of data as normal ACLs are too coarse-grained.} \\
        \hline
Decision Speed & Human Speed & Machine Speed

\emph{Increases velocity of checks because attackers are starting to operate
this way and we need to match their speed.} \\
\hline
Policy Type & Static (Allow/ Deny) & Static and Dynamic (Infer \& Interrupt)

\emph{Static policies are insufficient to codify the dynamic nature of today's
enterprises.}\\
\hline
Context Attributes & Identity + Device & Identity + Device + Behavior + Data Context 

\emph{A holistic view of user behavior matching or surpassing the capabilities
of a human analyst.}\\
\hline
Intent (User and Policy) & Simple Static Evaluation & User Intent + Agent Intent can be interpreted and checked to ensure
alignment and limit prompt injection risks. 

User Intent + Agent Intent can be an input in dynamic evaluation against
Policy Intent.\\
\hline
Action Window & Action only & Activity Window (before + after Action) 

\emph{Looking at an action in the context of the many operations that a user
or agent has performed both before and after their action can give a
more holistic picture of whether the action is legitimate.}\\
\hline 
Investigation & After-the-fact & Near-real-time and integrated back into authorization.

\emph{AI allows us to run rapid investigations in minutes instead of days or
hours, and act on the results of these investigations.}
\\
\hline
Challenges and Containments & Unlinked to Policy & Linked to Policy (e.g. Risk of Hijacking, trigger Security Key Touch) 

\emph{Reduced disruption to legitimate accessors compared to traditional
approvals, tailored hard challenges for malicious actors.}
\end{tabular}
\end{table*}

Implementing all the \emph{Beyond Zero} features allows an enterprise to express
high-level security policies that naturally follow resources across the
company and trickle down into individual enforcement actions.
Enterprises can also build both static and dynamic policies that take in
large amounts of data to make highly granular decisions, while
minimizing the degree to which human users are interrupted during their
workday. This unblocks a variety of use cases including access abuse,
intellectual property protection, and compliance use cases. For
accessors (humans and agents), the `access bubble' that controls their
permissions dynamically flexes to be bigger or smaller depending on what
they need to do in a given moment, reducing overprovisioning without
impacting productivity.

\section{Beyond Zero Architecture}\label{Beyond Zero-architecture}

\emph{Beyond Zero} operates by applying policy decisions to individual resources
rather than applications as a whole. Authorization is evaluated for a
specific resource, rather than a tool or application. Policies are
dynamically derived from reasoning decisions and immutable principles
that govern access. When an action is performed on data, a set of checks
are performed, taking into account various pieces of information about
the accessor, the data being accessed, and the operation being
performed, in order to authorize the access.

We maintain static policies (the "floor") to ensure baseline security
and compliance. Layered on top is a dynamic reasoning engine (the
"ceiling") that observes behavior and applies friction (challenges) when
an action deviates from established norms or carries high risk. There's
a statically enforced baseline that makes sure that the agent doesn't
deviate from a certain set of rules, then there's a dynamic component
that allows for decisionmaking at a higher granularity.

The Beyond Zero architecture consists of four cooperating components that
form a continuous feedback loop.

\subsection{\texorpdfstring{Autonomous Governance: Building the
Enterprise Security World Model
}{Autonomous Governance: Building the Enterprise Security World Model }}\label{autonomous-governance-building-the-enterprise-world-model}

Before a decision can be made, the system must understand the terrain.
In the same way that self-driving cars rely on a World Model to
understand the physics and constraints of the world in which they
operate, so too do we need to form an accurate Enterprise Security World Model to
help decide how agents and humans can take actions on resources
throughout the enterprise. Through the integration with existing HR and
project management warehouses and other sources of data that show who
humans inside the company are, and how their work interrelates to the
company's assets, the Autonomous Governance component uses AI to
generate and maintain a live Enterprise Security World Model:

\begin{itemize}
\item
  Who: Information about accessors (humans and agents), including their
  job function, role, seniority, controlling human (in the case of
  AI agents), etc. This information can be sourced from a variety of
  identity, HR, and other system of record databases. Preprocessing
  ensures that unstructured data is converted into static attributes
  that can be used as part of authorization. For example, where a person
  works in a given team, the subject matter of the team together with
  graph relationships of the team to other teams can be extracted from
  the organizational chart and other HR data, creating a set of
  attributes that can be used as part of security checks in both static
  and dynamic policies.
\item
  What: A semantic understanding of data sensitivity (confidentiality
  (e.g. confidential, crown jewels, etc), data type (e.g. user data)).
  This information is pre-processed by AI, allowing for resource
  understanding to be translated into an understandable structure. This
  could take the form of a semantic graph or a resource model (e.g.
  showing relationships between enterprise resources).
\item
  How: Information about what accessors are tasked to work on and how
  they do their work, based on historical usage and sharing patterns,
  group membership, team membership, assignments, etc. Preprocessing
  extracts the precise data around work assignments, for example that a
  user is currently working on an assignment that requires access to
  data belonging to Customers A, B and C only.
\end{itemize}

The need to make access decisions without adding undue latency to user
requests means that much of the information needed to make access
decisions must be accurately pre-populated and available at low-latency
at access time.

Pre-processing becomes critical because the strict latency budgets
available at access time mean that inference tasks must be front-loaded
to ensure that accurate attributes are available at access time.

The underlying principle here is to minimize the overhead of individual
real-time access decisions by pre-computing as much of the reasoning
outputs ahead of time. This pre-computed knowledge allows us to make
high-fidelity determinations with precision that wasn't previously
attainable for any conventional access control tools.

\subsection{\texorpdfstring{Event Intake: Connecting to Agent and
Human Activity
}{Event Intake: Connecting to Agent and Human Activity }}\label{event-intake-connecting-to-agent-and-human-activity}

To reason about risk, the system needs high-fidelity visibility not only
into the immediate action that's under evaluation, but also into the
broader context of user actions and enterprise patterns. To accomplish
this, \emph{Beyond Zero} can ingest streams of events from a variety of
traditional, existing data sources, including:

\begin{itemize}
\item
  Server-side: Access logs from corporate proxies and APIs. Enterprise
  productivity apps, system-level server access, and other server-side
  sources. Some of these signals can be used at-access-time without
  event enrichment, others require pre-processing to formulate
  attributes that are in turn used at access-time, or otherwise can be
  used in for post-access analysis for post-access actions.
\item
  Client-side: On-device signals indicating browser state, local file
  access, or process activity. This includes signals coming from
  standard DLP solutions and existing client security agents running on
  the endpoint, as well as raw system logs. These signals cannot be used
  without pre-processing, and typically are not useful other than in
  limited use cases, for example post-access.
\item
  Agent Activity: Logs of prompt inputs, execution plans, tool
  invocations, and more by AI agents. For agentic access scenarios,
  these additional signals can be used to determine whether the agent's
  access reflects both the intention of the user as well as alignment
  with enterprise policy.
\end{itemize}

The vast majority of this information is not directly acted upon by the
agent, but is used to gradually build a better understanding of user
behavior and to intercept actions once a certain risk threshold is
exceeded.

The information is stored in both a hot cache populated with processed
data for use in fast access-time policy evaluations, as well as a larger
long-term datastore that stores the larger set of data for more detailed
evaluation and pre-processing by different AI agents.

\subsection{Reasoning Engine: Making Static and Dynamic
Decisions}\label{reasoning-engine-making-static-and-dynamic-decisions}

This is the heart of \emph{Beyond Zero}. It is a hierarchical AI system
distributed across central server-side components and endpoint agents
that consumes both information about accessors, as well as recent access
signals to answer a single question: \emph{"In the context of what we
know about the user and the resource, is this specific action safe?".}
The reasoning engine in a similar way to the data storage systems is
capable of both fast reasoning tailored for at-access-time evaluation,
as well as slower inference to do richer analysis on a set of actions.

\begin{itemize}
\item
  Policy Evaluation (Fast) or Inference (Slow):

  \begin{itemize}
  \item
    In faster configurations, this can perform very granular
    attribute-based-access-control to determine things like a person
    accessing data belonging to a VIP user in one of the customer
    support tools, or a person attempting a high risk operation. These
    can be blocked at-access-time.
  \item
    In slower configurations, more complex anomalies can be detected,
    such as a human user accessing 500\% more files than their peer
    group. These can be then intervened with custom challenges or
    containments.
  \end{itemize}
\item
  Decision: It outputs a verdict: Allow, Deny, or Challenge. These
  decisions then factor into out-of-band risk evaluations which then
  become attributes that are referenceable in subsequent evaluations
  (even across attempts to access other applications or resources).
\end{itemize}

\emph{Attacker scenarios}

\emph{The curious contractor}

An external party attempts to compromise an enterprise by purchasing
stolen credentials of one of their contractors, who is working on
software development. As soon as the external attacker attempts to use
the credentials to start their attack, they are seen attempting to
access data not linked to the contractor's scope of work, and various
challenges can be issued to the accessor to confirm that the access is
legitimate.

In this scenario various aspects of the contractor's status are stored
as risk attributes, and the contractor's position in the organizational
chart along with their typical scope of work is stored as a set of
attributes linked to the accessor's identity. The autonomous governance
system has also performed analysis on the documents, understanding their
semantic context.

At access time, the reasoning engine sees that a high risk accessor with
low subject-matter crossover is attempting to access sensitive
documents. The contractor is given the option of requesting access via a
document owner approval challenge (see Challenge Infrastructure below on
how this works).

\emph{The suddenly foolish administrator}

In this scenario a system administrator logs into a service as usual,
but in another window, is looking up basic questions about the
architecture of the system that would be obvious to anyone experienced
at the company. These signals are stored in long-term storage, but are
then processed into an attribute indicating potential risk.

The administrator then attempts a host of operations, all of which
appear in a rapid succession indicating usage of agentic assistance. The
short term Reasoning Engine sees the risk indicator attribute and the
high risk operation attribute and stops the operation, challenging the
user to touch their security key to confirm that the user possesses both
the security key as well as their password, preventing compromise. After
clearing the challenge the record is stored in long term storage for
evaluation again should an unusual operation be seen again.

\emph{Automated reasoning}

As \emph{Beyond Zero} implementations mature, we also envision that the
reasoning engine could be used to reduce the burden of certain security
controls by deciding that users don't need to be challenged in scenarios
that are deemed low risk. Example: Security key touches might be seen as
unnecessary if an accessor is the same person showing up at the same
hour in the same location, but could then be re-applied should behavior
start to deviate from the expected baseline.

\subsection{\texorpdfstring{Challenge Infrastructure: Managing
Exceptions
}{Challenge Infrastructure: Managing Exceptions }}\label{challenge-infrastructure-managing-exceptions}

Challenge infrastructure is used to either gather context using
challenges, or restrict access using containments. Challenges are
introduced to gather further context before deciding whether to grant
access. Containments are more durable `stop signs' that are designed to
introduce substantial friction to an accessor's access to stop attacks.
Containments are designed to be dynamic and move up or down in strength
should a completed challenge clarify that the detected risk was a false
positive.

Challenges

When the Reasoning Engine detects ambiguity or elevated risk, it
triggers a "Challenge." Unlike the blunt "Access Denied" of the past,
challenges are granular and context-aware:

\begin{itemize}
\item
  Justification: "Please explain why you need access to this file."
\item
  Verification: "Touch your security key to prove you are present."
\item
  Approval: "Requesting manager approval for bulk export."
\item
  Biometric: "Please perform a selfie check to prove you're sitting at
  your machine."
\end{itemize}

Challenges are designed to be statically verifiable using policy,
however some challenges might contain dynamic evaluations that can be
then statically evaluated. For instance, policy can be written requiring
a check of a selfie before an administrator is allowed to perform a high
risk action, to ensure that they are sitting at their computer when the
action is being performed (and have not lost or sold their credentials).

More dynamic checks are also possible. For example, a person attempting
to email data to their personal account might be asked to provide a
justification, and this justification can be checked against the
document contents in real time.

The goal of the Challenge Infrastructure is to abstract away the
complexity of implementing each of these challenges into a single
service that can have static and dynamic policies written against it.
That decoupling ensures that it can independently evolve as we have
better challenges available, for example should a company roll out a
fingerprint-enabled security key, this could easily be rolled out as a
challenge for the system to use.

Containments

Containments are a second, more durable type of interruption. These
interruptions revoke some or all of the actor's access in response to a
perceived risk. Containment revocation is generally not as
straightforward as challenge completion. While some containments can be
revoked by correctly passing a challenge, other containments might only
be lifted by the security team who might interview the contained user
and their manager before deciding on a course of action.

\section{An End-to-End Example: The "Rogue"
Agent}\label{an-end-to-end-example-the-rogue-agent}

Consider a scenario where an internal AI agent, SalesGenie, is
authorized to read sales reports.

\begin{enumerate}
\def\labelenumi{\arabic{enumi}.}
\item
  Action: SalesGenie attempts to query a highly sensitive strategic
  planning document inside Google Drive.
\item
  BeyondCorp Check: Both the agent and the person who prompted the agent
  have a valid certificate from an authorized machine and identity, and
  both identities are authorized to access sales reports. Access
  Allowed.
\item
  \emph{Beyond Zero} Check:

  \begin{itemize}
  \item
    Accessor Context: The Reasoning Engine sees that the user that
    prompted SalesGenie only works on Financial Services accounts in the
    Northeast. The user's risk score is also higher as they have been
    recently attempting accesses that are not closely related to their
    work assignment.
  \item
    Data Context: The data being accessed is highly strategic, company
    level data only used in financial reporting. The only teams that
    typically access this data are the Strategy and Finance teams. The
    data is classified as Highly Confidential, the highest tier of
    sensitivity, as it could be used in insider trading.
  \item
    Decision: The Reasoning Engine finds a policy specifying that for
    data classed as Highly Confidential, a valid work assignment must be
    present. As the work assignment here does not match the worker types
    who need access to this data, the accessor is served two challenges.
    A second, slower inference can happen over the actor's recent
    accesses once the first risky action has been detected. This can
    lead to intervention once the detailed analysis has been performed
    by the automated reasoning engine. In this case, the inference might
    detect a pattern of accesses together with logs showing the actor
    attempting to exfiltrate some of their data, and decide that a
    stronger containment is needed.
  \item
    Intervention:

    \begin{enumerate}
    \def\labelenumii{\roman{enumii}.}
    \item
      The human actor must confirm that the access the agent is
      attempting is intended by the human, to minimize the risk of rogue
      actions by the agent
    \item
      The human actor must then request approval from the team that owns
      the data confirming that they require access to it
    \item
      Following the failure to clear challenges and a more detailed
      inference completing over a larger set of actions, the user is
      then contained. In the vast majority of cases, the decision to
      contain will be autonomous. Only a tiny percent would be escalated
      to human review when we have a high confidence in malicious
      intent, which is substantially higher fidelity than the current
      paradigm.
    \end{enumerate}
  \end{itemize}
\end{enumerate}

\section{Accelerating towards Beyond Zero in the technology
industry}\label{accelerating-towards-Beyond Zero-in-the-technology-industry}

In our view a shift towards a \emph{Beyond Zero} model will require landing a
number of major alignments across the industry more broadly. These would
be useful areas for exploration:

1) Open architectures that can enhance transparency into access as
enterprises make the agentic transformation. For example, standardized
APIs for agent introspection, with a standardized way to analyze
chain-of-thought / tool use in real time.

2) Standards for agentic identities and how to define and control their
access. For example, an industry standard for request annotations to
make all agent actions to be attributable to a specific agent, a
specific controlling user, and a specific task.

3) Frameworks for allowing reasoning engines to safely make access
decisions and take actions. For example, make external, enterprise
operated policy evaluation and decision point integration and
enforcement a first-class citizen for all SaaS services.

NIST has already commenced an Agent Security effort, and we're hopeful
that efforts like this will accelerate the industry's path towards
making these important standardizations.

\section{Conclusion}\label{conclusion}

The `Application Boundary' model is reaching its end of life as the rise
of autonomous agents and exponential data velocity stretch legacy Zero
Trust frameworks to their breaking point. As AI agents join our
workforce and the speed of business accelerates, we must adopt a
security model that works at machine speed as well.

\emph{Beyond Zero} represents the next major evolution in enterprise defense,
shifting the trust boundary from the broad application level to the
individual action performed on a specific resource. By unifying static
authorization with a dynamic, AI-driven reasoning engine, the
architecture enables a self-defending enterprise capable of mediating
thousands of decisions per second at machine speed. It effectively fuses
access management and security operations into a single, high-frequency
feedback loop that can infer and interrupt threats in real-time.

We believe \emph{Beyond Zero} is the future of enterprise security, moving
towards a model of security as an immune system that continuously adapts
to the context and intent of every request. This transition is a
strategic necessity for the agentic era, ensuring that enterprise data
remains secure even as the boundaries between human and machine actions
continue to blur. Our hope is that these ideas will help make the
industry safer, but also that others will build on this foundation to
strengthen the state of the art of how security should work in this new
era. Now that this future is coming into view, it's time to accelerate
our investments towards the development of new standards, particularly
around pluggable policy evaluations and agent context annotations.

As we continue our own journey in developing and deploying \emph{Beyond Zero},
we intend to continue publishing more papers in this space. We encourage
others to start the same journey and share their learnings with the
industry as we all work through the challenges of this new era together.

\section*{Author's Note}

\emph{The Beyond Zero approach is the culmination of years of thinking
and work by many individuals at Alphabet and in the broader industry. In
alphabetical order we'd like to thank:}

\emph{
Heather Adkins,
Darren Bilby, 
Gordon Chaffee,
Garrett Cronin, 
Ian Green,
Royal Hansen,
Sebastian Harris,
Gregory Kick,
Kaan Kivilcim,
Bryan Landsiedel,
Naveed Makhani,
Justin McWilliams,
Pankaj Rohatgi,
Ismail Sebe,
Ruchi Shah, 
Umesh Shankar,
Nic Shupe, 
Adam Tanana,
Marianna Tishchenko,
Tavish Vaidya,
Joseph Valente,
Juan Vasquez, 
Ollie Wild,
Michal Zalewski}

\section*{References}
BeyondCorp: A New Approach to Enterprise Security, Rory Ward, Betsy Beyer, ;login:, Vol. 39, No. 6 (2014), pp. 6-11.
\end{document}